\documentstyle[aps, preprint, psfig]{revtex}
\begin{document}
\draft
\title{Effect of defects on thermal denaturation of DNA oligomers}
\author{Navin Singh and Yashwant Singh}
\address{ Department of Physics, Banaras Hindu University, \\
     Varanasi 221 005, India } 
\maketitle

\begin{abstract}
The effect of defects on the melting profile of short 
heterogeneous DNA chains are calculated using the Peyrard-Bishop 
Hamiltonian. The on-site potential on a defect site
is represented by a potential which has only the short-range
repulsion and the flat part without well of the Morse potential. 
The stacking energy between the two neigbouring pairs involving 
a defect site is also modified.
The results are found to be in good agreement with the experiments.
\end{abstract}
\narrowtext
\vspace{0.5cm}

 DNA thermal denaturation is the process of separating the two strands 
wound in a double helix into two single strands upon heating. 
Several experiments \cite{1,2} on dilute DNA solutions have provided
evidence for the existence of a thermally driven
melting transition corresponding to the sudden opening
of base pairs at a critical(or melting) temperature $T_m$.
At  $T_m$ one-half of the DNA denatured. 
The understanding of this remarkable one-dimensional
cooperative phenomena in terms of standard statistical
mechanics, {\it i.e.}, a Hamiltonian model with temperature
independent parameters is a subject of current interest \cite{3,4,5,6,7,8}.
 
 DNA is modelled as a quasi one dimensional
lattice composed of N base pair units \cite{9}. 
The double helix is held together by hydrogen bonds
between complementary bases on opposite strands and
hydrophobic or stacking interactions between nearest 
neighbor bases on opposite strands. Each base pair is 
in one of the two states; either open (non-hydrogen) or
intact (hydrogen-bonded). A Hamiltonian model which has been
found appropriate to represent  interactions in DNA is the 
Peyrard-Bishop (PB) model \cite{3,4}. The PB model is written as,

\begin{equation}
H = \sum_n[{\frac{p_n^2}{2m} + W(y_n, y_{n+1}) + V_n(y_n)}] 
\end{equation}
where $m$ is the reduced mass of a base pair, $y_n$ denotes
the stretching of the hydrogen bonds connecting the two 
bases of the $n^{th}$ pair and 

 $$p_n = m(\frac{dy_n}{dt})$$

 The on site potential $V_n(y_n)$ describes the interaction of the
two bases of the $n^{th}$ pair. The Morse potential,

\begin{equation}
V_n(y_n) = D_n(e^{-ay_n} - 1)^2
\end{equation}
which is usually taken to represent the on-site interaction represents
not only the H bonds connecting two bases belonging to opposite 
strands, but also the repulsive interactions of the phosphates 
and the surrounding solvents effects. The flat part 
at large values of the displacement of this potential emulates
the tendency of the pair "melt" at high temperatures as thermal
phonons drive the nucleotides outside the well (see Fig.1) and
towards the flat portion of the potential.  

The stacking energy between the two neighboring base pairs is 
described by the anharmonic potential

\begin{equation}
W(y_n, y_{n+1}) = \frac{k}{2}[1 + 
                      \rho e^{-\alpha(y_n + y_{n+1})}]
                        {(y_n - y_{n-1})}^2 
\end{equation}

 The choice of this form of $W(y_n, y_{n-1})$ has been
motivated by the observation that the stacking energy is not
a property of individual bases but a character of the base pair,
themselves \cite{7}. When due to  stretching the hydrogen bonds 
connecting the bases break, the electronic distribution on bases
is modified causing the stacking interaction with adjacent bases 
to decrease. This is taken into account by the exponential term 
in Eq.(3).  One may note that the effective coupling 
constant decreases from $k(1+\rho)$ to $k$ when either one of
the two interacting base pairs is stretched. Eq.(3), therefore,
takes care of changes in stacking energy due to breaking of hydrogen
bonds in base pairs due to stretching. This decrease in 
coupling provides a large entropy in the denaturation. The parameter
$\alpha$ in Eqs.(3) defines the "anharmonic range". 

 The model Hamiltonian of Eq.(1) has extensively been used to 
study the melting profile of a very long $(N\to\infty)$
and homogeneous DNA chain using both statistical mechanical
calculations and the constrained temperature molecular dynamics
\cite{10,11}. Analytical investigation of nonlinear dynamics of the
model suggests that intrinsic energy localization can initiate the
denaturation \cite{12}. In case of long homogeneous DNA chain the
model exhibits a peculiar type of first-order transition with
finite "melting entropy", a discontinuity in the fraction of 
bound pairs and divergent correlation lengths. However, as the 
value of the stacking parameter $\alpha$ increases and the 
range of the "entropy barrier" becomes shorter than or comparable
to the range of the Morse potential the transition changes to 
second order. The crossover is seen at 
$\alpha/a = 0.5$ \cite{8}.

 When one considers chain having inhomogeneity in base pair 
sequence, one faces a problem in investigating its statistical
mechanics as the transfer-integral method which has been used
in the case of homogeneous chain is no longer valid. Attempts
have, however, been made to use the model Hamiltonian of Eq.(1)
for heterogeneous chains either by modelling the heterogeneity
with quenched disorder \cite{6} or by properly choosing basis
sets of orthonormal functions for the kernels appearing in the
expansion of the partition function \cite{7}. 

 In this note we investigate the effect of defects on the
melting profile of short DNA chains of heterogeneous 
compositions. A defect on DNA chain means a mismatched base-pair.
For example, if one strand of DNA has adenine on a site
the other strand has guanine or cytosine instead of thymine
on the same site. In such a situation the pair will remain
in open state at all temperatures as two nucleotides cannot
join each other through hydrogen bonds. Oligonucleotide probes
are commonly used to identify the presence of unrelated 
nucleic acids. In this context it is therefore important 
to discriminate the targets that differ from one another
by a single or more nucleotides. 

 Since the partition function of the PB model is convergent 
only in the limit of infinite number of base-pairs N 
({\it i.e.} $N\to\infty$) \cite{4}, 
it is therefore, necessary in the case of short chains to 
consider not only the breaking of hydrogen bonds between 
single base pair, but also the complete dissociation of the two 
strands forming the double helix. In other words, in the
case of short chains there is a need to consider thermal
equilibrium between dissociated strands and associated double
strands(the duplex) and a thermal equilibrium in the duplexes,
between broken and unbroken interbase hydrogen bonds.

 The average fraction $\theta$ of bonded base pairs can be
factored as $ \theta = \theta_{ext}\theta_{int} $ \cite{7,13}.
$\theta_{ext}$ is the average fraction of strands forming duplexes,
while $\theta_{int}$ is the average fraction of unbroken bonds 
in the duplexes. The
equilibrium dissociation of the duplex $C_2$ to single strand
$C_1$ may be represented by the relation $C_2 \rightleftharpoons 2C_1$
\cite{8,9}.  The dissociation equilibrium can be neglected in the case
of long chains; while $\theta_{int}$ and thus $\theta$ go to
zero when $\theta_{ext}$ is still practically 1. This is 
because in the case of long DNA fragments when $\theta$ goes
practically from 1 to 0 at the melting transition, the two
strands may not get completely separated; while most bonds
are disrupted and the DNA has denatured, few bonds still 
remaining prevent the two strands from going apart each other.
It is only at $T>>T_m$ there will be a real separation. Therefore,
at the transition the double strand is always a single molecule
and in calculation based on the PB model one has to calculate
only $\theta_{int}$. On the contrary,
in the case of short chains the process of single bond disruption
and strand dissociation tend to happen in the same temperature
range; therefore, the computation of both $\theta_{int}$ and 
$\theta_{ext}$ is essential.

 For the computation of $\theta_{int}$ one has to separate
the configurations describing a double strand on the one hand, and
dissociated single strand on the other. For this we follow
the method suggested by Campa and Giansanti \cite{13}.  
The $n^{th}$ bond is considered  
open if the value of $y_n$ is larger than a chosen threshold
$y_0$. A configuration belongs to the double strands if at least
one of the $y_n^{'s}$ is smaller than $y_0$. One can therefore
define $\theta_{int}$ for an N base pair duplexes by:

\begin{equation}
\theta_{int} = \frac{1}{N} \sum_{i=1}^{i=N}
                    \langle \vartheta(y_0 - y_i)\rangle
\end{equation}

 where $\vartheta(y)$ is Heaviside step function and the
canonical average $\langle.\rangle$ is defined considering only the double
strand configurations. For  $y_0$, we have taken a value of 2 $\AA$.
Since the PB model couples only the nearest neighbors the 
calculation of canonical averages in Eqs.(4) reduced to 
multiplication of finite matrices. The discretization of the
coordinate variables and introduction of a proper cutoff on the maximum
value of $y^{'s}$ \cite{10} determines the size of the matrices
and the number of base pairs in the chain the number of matrices
to be multiplied.

 Since we are concerned with a heterogenous DNA chain, we have 
selected two different values of $D_n$ (see Eq. 2) according to 
the two possible base pairs; adenine-thymine (A-T) and 
guanine-cytosine(G-C). While A-T has the two hydrogen bonds, the 
G-C pair has three. Because of this $D_n$ for G-C bonds is chosen 
as nearly 1.5 times larger than the one representing 
to the  A-T bonds. The complete set of values in our calculations
are:$D_{AT}  =  0.05 \;\; eV , \;\; D_{GC}  =  0.075 \;\; eV,  
\;\; a_{AT}  =  4.2 \;\;{\AA^{-1}}, \;\; a_{GC}  =  6.9,  \;\;{\AA^{-1}}   
\;\; k  =  0.025 \;\; eV/{\AA^2}, \;\; \rho  =  2, \;\; 
{\rm and} \;\; \alpha  =  0.35 \;\; {\AA^{-1}}$

Since at a defect site the nucleotides of the two strands do not
associate themselves through hydrogen bonds we replace  the on-site
Morse potential by a potential shown in Fig.1  by full line. 

\vspace{0.5cm}
\psfig{figure=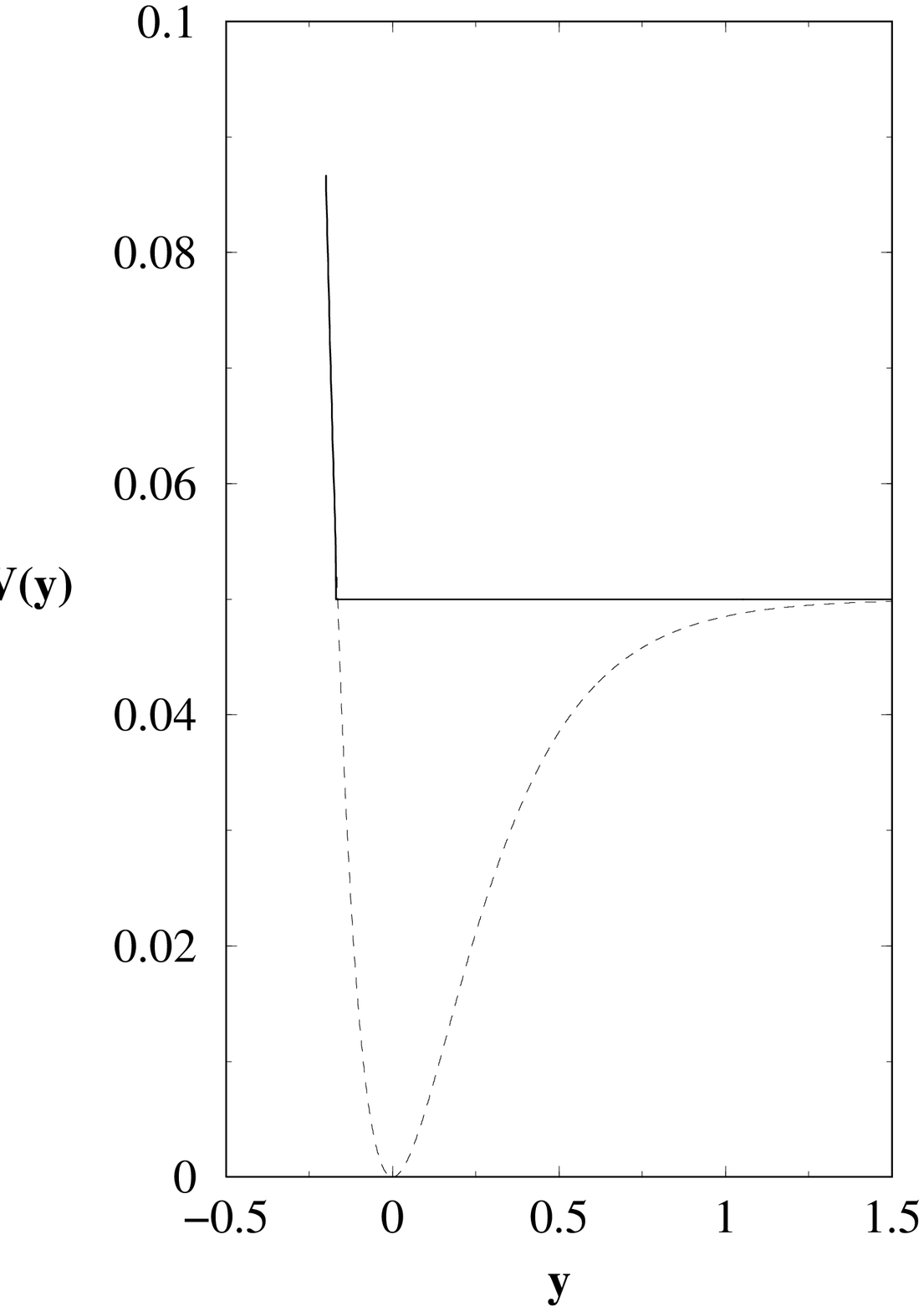,height=2.5in,width=2.5in}
{Fig. 1 \small The on-site potential V(y) as a function of 
displacements. The dotted line represents the Morse potential
(see Eqn.(2) and the full line the potential chosen to represent
the interaction at the defect sites.}
\vspace{0.5cm}

This potential has repulsive part as well as the flat part of the
Morse potential but not the well which arises due to hydrogen
bonding interactions.
Due to defect on a site the stacking interactions with adjacent bases 
will also be affected. As argued above, the formation of hydrogen bonds
changes the electronic distribution on base pairs causing stronger
stacking interactions with adjacent bases. Therefore, when one of
the base pairs without hydrogen bonds is involved the stacking 
interaction will be weaker compared to the case when both base pairs 
are in intact state. This fact has been taken into account 
in our calculation by reducing the anharmonicity coefficient $\rho$
to  its half value wherever the defective site appeared in Eq.(3).

 For $\theta_{ext}$ we use the expression given in Ref. [13]. Thus,

\begin{equation}
 \theta_{ext} = 1 + \delta - \sqrt{\delta^2 + 2\delta}
\end{equation}

where
\begin{equation}
  \delta = \frac{Z(A)Z(B)}{2N_0 Z(AB)} 
\end{equation}
  
 Here Z(A), Z(B) and Z(AB) are the configurational isothermal-
isobaric partition functions of systems consisting of molecular
species A, B and AB respectively. $N_j$ is the number of molecules
of species $j$ in volume V and $2N_0 = 2N_{AB} + N_A + N_B$.
In deriving Eq.(6) $N_A$ has been taken equal to $N_B$ and
$\theta_{ext}$ defined as 

$$ \theta_{ext} = \frac{N_{AB}}{N_0} $$

 The partition function for each species can
be factored into internal and external part and can
be written as \cite{13},

\begin{equation}
 \frac{Z(A)Z(B)}{2N_0 Z(AB)} =  
    \frac{Z_{int}(A)Z_{int}(B)}{a_{av}Z_{int}(AB)} 
 \frac{a_{av}Z_{ext}(A)Z_{ext}(B)}{2N_0 Z_{ext}(AB)} 
\end{equation}
  
 where $a_{av.} = \sqrt{a_{AT}a_{GC}}$

In analogy to what has been proposed for the Ising model \cite{3}
on the basis of partition function of rigid molecules \cite{16}, one makes
the following choice 
 
\begin{equation}
 \frac{a_{av.}Z_{ext}(A)Z_{ext}(B)}{2N_0 Z_{ext}(AB)} 
  = \frac{n^*}{n_0}N^{-p\theta_{int} + q} 
\end{equation}

where $n^*$ is a chosen reference concentration as 
1$\mu M$ while $n_0$ is the single strand concentration
which we have chosen as 3.1 $\mu M$. $p$ and $q$ are the
parameters which can be calculated using experimental
results. 

 We have considered the following two different oligonucleotides
with sequence given by:

\begin{itemize}
\item [({\bf A})] $ ^{5^{\prime}} GTGTTAACGTGAGTATAGCGT_{3^{\prime}}$
\item [({\bf B})] $ ^{5^{\prime}} GGT_{11}GG_{3^{\prime}} $
\end{itemize}

and by the respective complementary strands when there is
no defect. The melting profile of oligonucleotides (A) has been calculated
by Campa and Giansanti \cite{13}. They found very good agreement
with experiment. We have extended their calculation and have 
studied the effect of defects on the melting profile. The 
results are plotted in Fig. (2). Our results for defectless
chain agree very well with that given in refs. \cite{13} and
with experiments \cite{14}. 

\vspace{0.5cm}
\psfig{figure=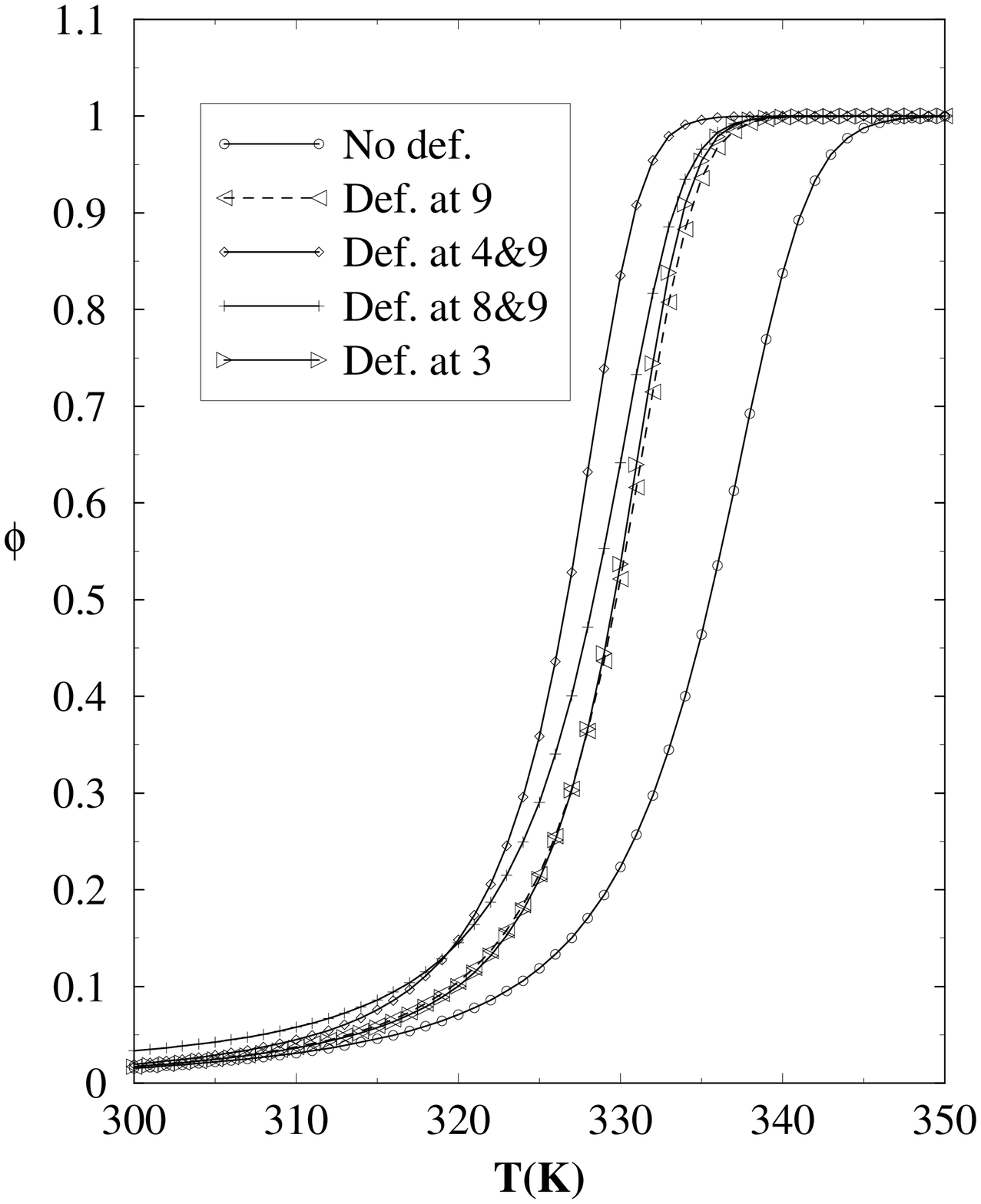,height=2.5in,width=2.5in}
{Fig. 2 \small Plot showing the variation of melting temperature with
defect. Here $p=29.49$ and $q=27.69$}
\vspace{0.5cm}

\vspace{0.5cm}
\psfig{figure=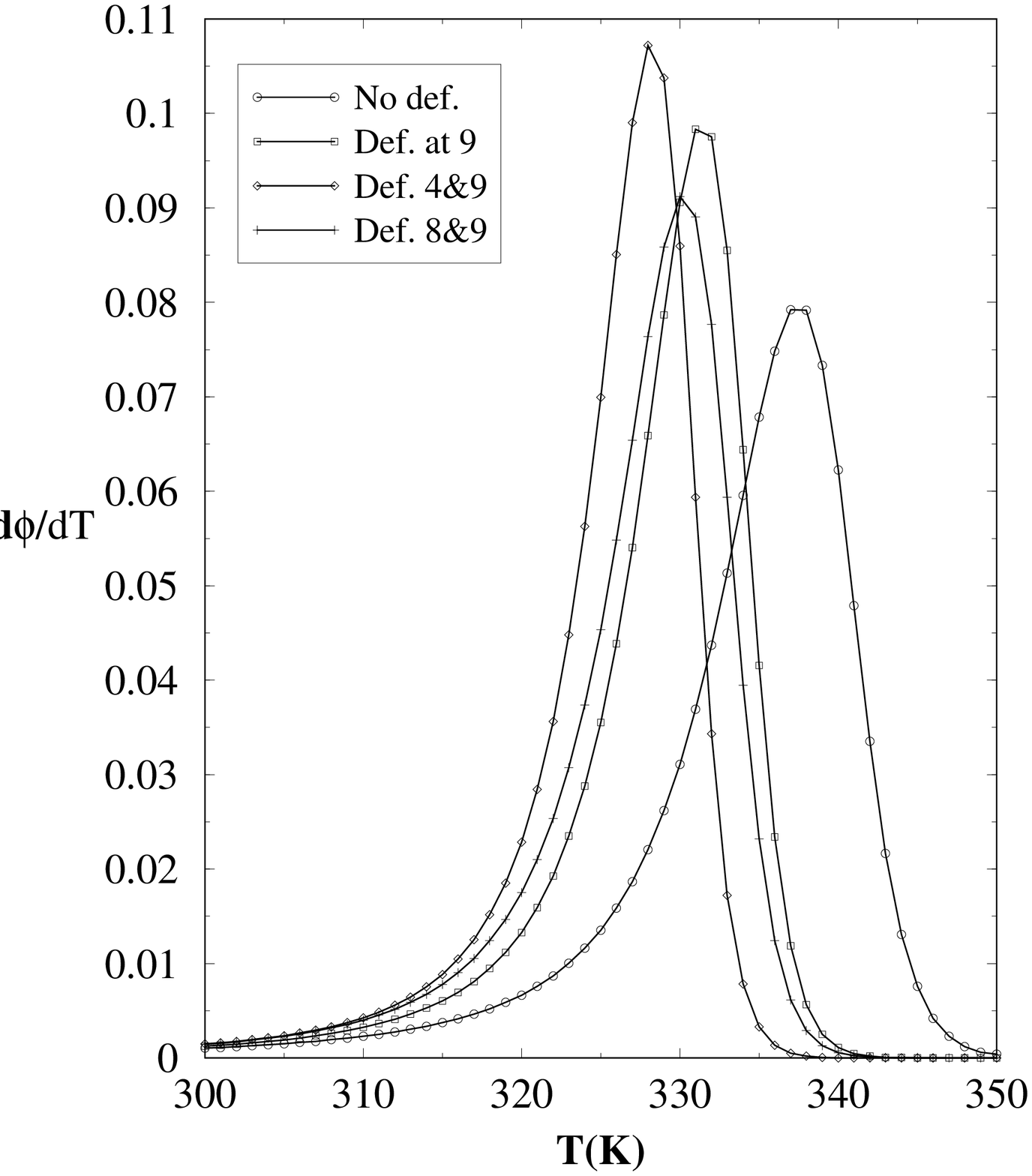,height=2.5in,width=2.5in}
{Fig.3 \small Plot showing the differential melting curve}
\vspace{0.5cm}

We used the same parameters except the
modification as stated above in the Morse potential and the anharmonic 
stacking potential term to calculate the melting profile with one 
defect on a site. Our calculation indicates that the melting 
profile of the chain does not depend on the location of the defect site.
When we introduced two defects one at $4^{th}$ and the other 
at $9^{th}$ sites we found the melting curve shifts to further
lower temperature. While we found the melting temperature
$T_m$ decreases from 335.7 K to nearly 329.7 K ({\it i.e.} 
$\Delta T_m^{(0,1)} \sim 6 K$) in the presence of one
defect, the decrease in $T_m$ from one to two defects is
only $\Delta T_m^{(1,2)} \sim 3 K$. These results are in good
agreement with the experimental observations \cite{14}.
We also put the two defects at the consecutive sites;{\it i.e.}, at 
$8^{th}$ and $9^{th}$ sites.  As Fig. 2 shows there is a change 
in the melting profile compared to the case when the two defect sites
were apart. This change is more clearly seen in the plot of 
$d\phi/dT$ shown in Fig.3.

 The oligonucleotides (B) has been studied by Bonnet {\it et al}
\cite{15}. They measured the change in the melting temperature
with one defect placed on different sites of the chain. They
found $T_m = 315 K$ for defectless chain and $\Delta T_m = 8 K$
with one defect. Their results shows that the location of the defect
on the chain has practically no effect on the melting temperature
of the chain in agreement with our calculations.

\vspace{0.5cm}
\psfig{figure=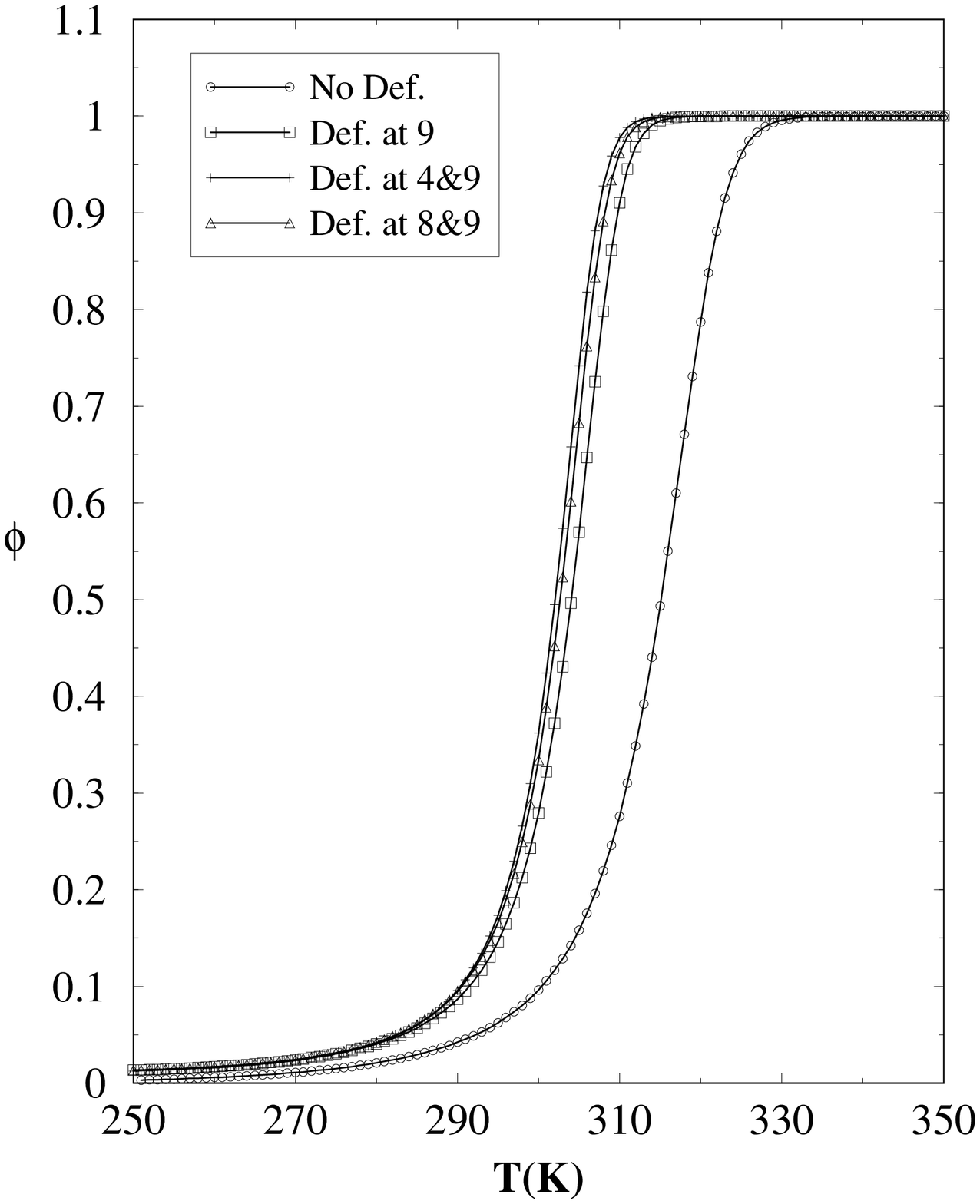,height=2.5in,width=2.5in}
{Fig. 4 \small Plot showing the variation of melting temperature
with defect. Here $p=34.46$ and $q=32.45$}
\vspace{0.5cm}

\vspace{0.5cm}
\psfig{figure=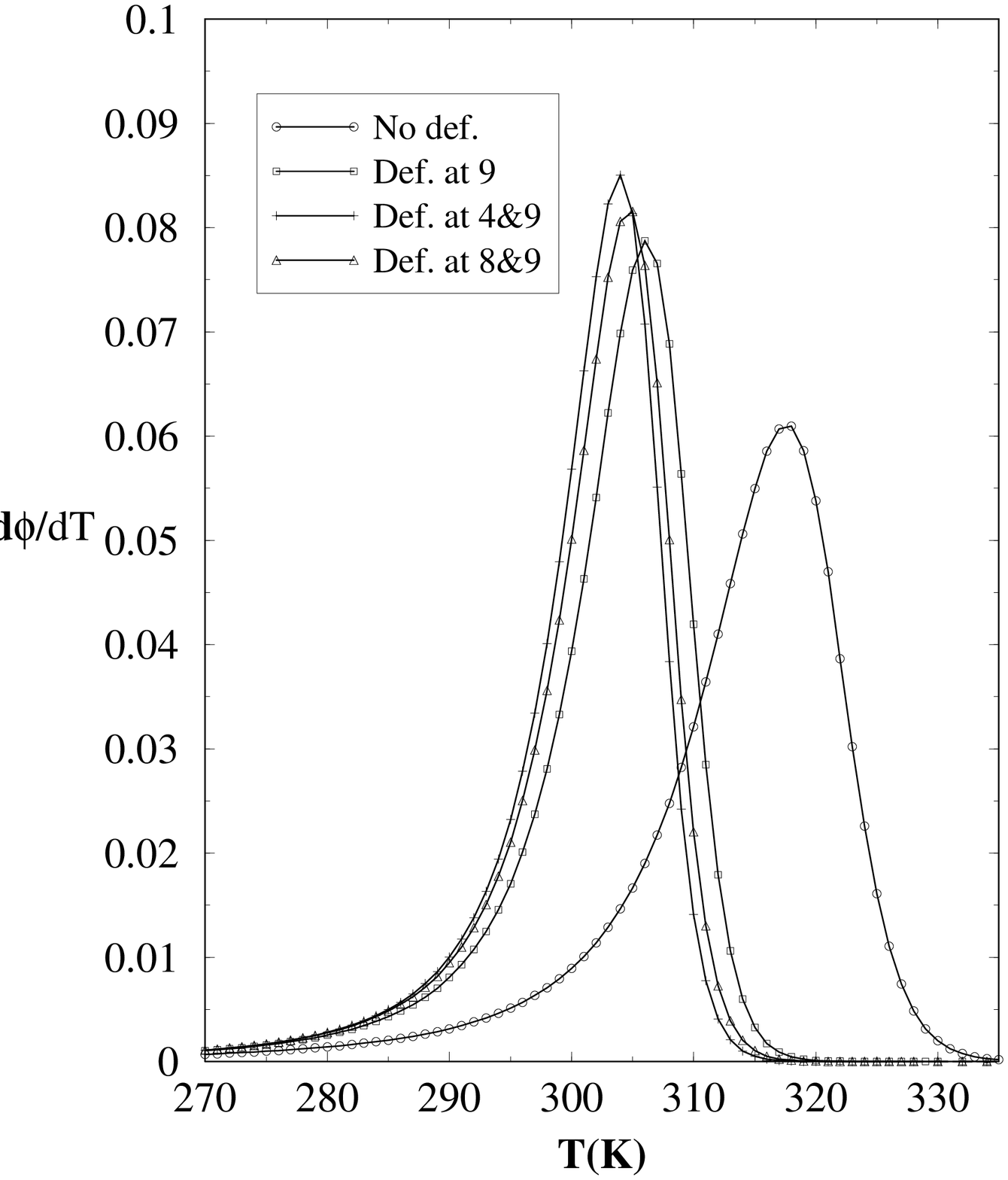,height=2.5in,width=2.5in}
{Fig. 5 \small Plot showing differential melting curves}
\vspace{0.5cm}

 We plot our results in Fig. 4. The force parameters are same as
given in Eq.(5). Since the experimental conditions are slightly
different than the one for chain (A) we adjusted parameters
$p$ and $q$ so that melting profile for the defect less chain
agrees with the experimental result. Using these values of the
parameters we have calculated the melting profile with one
and two defects. We find that while  $\Delta T_m^{(0,1)} \sim 10 K$
and $\Delta T_m^{(1,2)} \sim 3 K$. Fig. 5 shows the plot of
$d\phi/dT$.

 In conclusion we wish to emphasize that our calculations show
that the PB model is capable of describing the melting profile
of oligonucleotides of heterogeneous composition in the presence
of defects also.   

 This work was supported by the Department of Science and 
Technology, (India) through research grant.

\end{document}